\documentclass{llncs}

\usepackage{todonotes}
\usepackage{graphicx}

\begin{document}

\title{Beyond Participatory Design: Towards a Model for Teaching Seniors Application Design}
%
%
\author{Dorota Orzeszek\inst{1} \and Wieslaw Kopec\inst{1} \and Marcin Wichrowski\inst{1} \and Radoslaw Nielek\inst{1} \and Bartlomiej Balcerzak\inst{1} \and Grzegorz Kowalik\inst{1} \and Malwina Puchalska-Kaminska\inst{2}}
\authorrunning{Dorota Orzeszek et al.} 
%
%
\institute{\textsuperscript{1}Polish-Japanese Academy of Information Technology, Warsaw 02-008, Poland,\\
\email{dorota.orzeszek@pja.edu.pl}\\
\textsuperscript{2}University of Social Sciences and Humanities, Warsaw 03-815, Poland}
\maketitle              

\begin{abstract}

Population aging and the ubiquity of technology in everyday life have made designing solutions for older adults a necessity. User-centered and participatory design approaches include elderly users in the software development process to some extent but do not encourage them to take a leading role in designing applications to address their unmet needs. Teaching seniors about software design could help them actively participate in creating much needed solutions for their age group but this cannot be done without first understanding their conceptual models of technology. Past experiences play a significant role in determining the way learners model abstract concepts and so older adults' conceptual models of user interfaces (and human-computer interaction in general) differ from those used in teaching application design to younger students. In this paper we analyze a workshop on user interface design and prototyping for seniors to better understand older adults' learning process and the issues they encounter while learning abstract ideas related to human-computer interaction. We conclude the study by proposing guidelines for teaching older adults abstract technology related concepts.

\end{abstract}

\section{Introduction}

According to the United Nation's population projections, by 2050 over 20\% of the world population shall be over 60 years old and the percentage of those over 80 shall double \cite{UNreport}. In the USA alone 87.3 million people over 65 are expected \cite{UScensusreport}. This trend and the fact that information and communication technology (ICT) has become an immanent part of everyday life means engaging seniors in developing technology solutions to suit their specific needs is becoming ever more important. One way of doing this is through participatory design where end users help design and test solutions. In our study we try to take participatory design a step further by engaging seniors in the whole application prototyping process and move from "designed for older adults" to "designed by older adults".

The primary goal of this study is to verify whether older adults with no previous design experience are able to comprehend abstract technology concepts such as interfaces and interactions, internalize the idea of user interface (UI) prototyping and build adequate conceptual models of applications and the software design process. The study identifies the difficulties seniors face when learning these abstract concepts and proposes ways to address them.

The results of this preliminary study shall be used to propose a complete methodology for successfully teaching older adults software and user experience (UX) design skills which would in turn allow them to engage more actively in participatory design initiatives and take the role of full-fledged design team members instead of merely consulting younger designers' ideas.

As modeling human perception of abstract concepts is a complex topic, the results of this study are on their own not enough to propose a fully developed software design teaching methodology, however, they do provide some insights into the issues faced by elderly learners in learning UX design. These insights may be a valuable hint for organizations working with seniors when introducing them to application design and other abstract technology concepts and may help avoid the most common issues that would otherwise hinder their learning.

The article begins with a review of the current literature in the areas of participatory design, conceptual modeling, teaching models for working with older adults and UX training principles (Section 2). In section 3 the course settings are outlined. Section 4 focuses on the methodology used in this study. Observations are noted and discussed in section 5. In the final section conclusions are drawn from the observations and suggestions for solutions are proposed. 

\section{Related Work}

\subsection{Participatory design}

The rise of user experience in the 1990s \cite{norman_design_1988} has shifted the industrial software design process from the long-lasting system-centered design paradigm to an approach which placed end users at its core \cite{johnson_use-centered}. The user-centered design paradigm has ever since been the prevalent approach to application design \cite{getto_teaching_2013}. This method utilizes the standard 4-phase software development process (analysis, design, prototyping, testing) but the testing phase is performed in cooperation with real end users from the software's target group \cite{ladner_design_2015}.

Scholars stress the social nature of software design \cite{wang_software} which has led to the creation of participatory design. As defined by \cite{ladner_design_2015}, participatory design builds on the user-centered approach but involves end users in both the testing and design phases. This allows users to influence the final product to a greater extent as they are partially involved in designing the solution. 

Once the influence of aging on the perception of user experience had been realized \cite{madeiros_influence}, numerous case studies have analyzed the process of engaging seniors in participatory design \cite{demirbilek_involving_1999}, 
evaluating conceptual designs with elderly end users \cite{demirbilek_universal_????}, \cite{demirbilek_collaborating_2000}, also in the context of mobile design \cite{nicol20152nd}. Scholars have also compared seniors' and designers' mental models of mobile devices to try to provide guidelines for product and UX designers \cite{tang_understanding}.

\subsection{Conceptual modeling in user interface design}

Psychology and conceptual modeling are key in designing intuitive user interfaces and providing a satisfying user experience. Software applications materialize a system's underlying conceptual model \cite{pastorBeyond} and convey it via their UI which should converge with the user’s mental model of the system. A conceptual model is not complete merely after specifying requirements for the system's functionalities - end users' interactions with the system also shape the underlying model \cite{aquinoConceptual}. A user-centered conceptual model for an application's user stories must be taken into account \cite{dusterhoftConceptual}. 

Users’ mental models of technology determine their understanding of and interactions with software systems but are influenced by individuals' knowledge, skills and past experiences \cite{dusterhoftConceptual}. Thus, specific user groups (e.g. older adults) cannot be expected to accept and adapt to conceptual models tailored to other groups.  

\subsection{Mental models and teaching older adults technology}

Teaching older adults to use ICT is a key concern for many governments \cite{naumanen_guiding_2007}. Studies have been conducted in order to determine ways of engaging older adults in ICT - from examining their first interactions with new devices \cite{burrows_together}, through determining their habits in using mobile devices \cite{levy2016keeping} and strategies while searching the Internet \cite{karanam2016age}, to studies on overcoming computer anxiety in seniors \cite{cooper2015computer}.

When preparing ICT curricula for older adults teachers need to compensate for seniors' cognitive decline and sensory deficits \cite{acharya_aging} in order for them not to lose their self confidence and learning motivation \cite{boultonlewis-ageing-2007}. 

Another important factor is older adults' different understanding of technology and the mental models they use when learning ICT. Studies point out many issues older adults face while learning to use modern technologies, i.a. not grasping different contexts and thus confusing similar UI elements (e.g. the search and address fields in a web browser) \cite{naumanen_guiding_2007}. Seniors are also more thorough in their searches for information, i.e. they are biased towards entering and examining more hyperlinks while younger users visually scan search results and click on only a few links \cite{haeggans_60s_2012}. Seniors are generally more prone to being distracted by irrelevant information and losing focus \cite{naumanen_practices_2008}. When combined with the age-related deterioration of working memory \cite{gamberini_cognition_????}, these cognitive predispositions may substantially influence seniors ICT learning abilities. 

\subsection{Teaching UX and prototyping}

The necessity of teaching UX skills has long been acknowledged by both industry and academia \cite{getto_teaching_2013}. The question of selecting an appropriate curriculum has been widely debated - both in the context of the pedagogical challenges of teaching user experience and emphatic design \cite{gagnon_learning} as well as selecting a comprehensive set of topics to train industry-ready UX designers \cite{getto_toward}.

The basis for a successful UX curriculum outlined by practitioners consist of i.a. an overview of core areas of UX (user research, information architecture and, interaction design), UX problem solving, emphasizing the role of the user at every stage of product design, explaining the balance between user needs, business goals and technical constraints \cite{ivins_what}. \cite{getto_teaching_2013} also stresses the importance of teaching conceptual understanding, skills in using appropriate tools and offering hands-on practice problems.

Most curricula are aimed at graduate level students and young professionals \cite{getto_teaching_2013}. To the best of the authors' knowledge, teaching UX to older adults is an area that has yet to be explored. In \cite{nicol20152nd} several key issues in designing with older adults are highlighted e.g. their lack of confidence that their ideas can be transformed into real products and not understanding the need for providing both paper and interactive prototypes to properly test a design. Teaching seniors the skills and vocabulary necessary to convey their needs and ideas to a development team has much potential to solve many of the problems of participatory design.

\subsection{Design for empowerment}

Studies on engaging elderly users in software development using the user-centered and participatory design approaches are not a novelty in the field - participatory design use case studies have been conducted since the late 90s \cite{demirbilek_involving_1999}, \cite{demirbilek_collaborating_2000}. One interesting case study engaging users not only in testing but also partially in designing features has been described in \cite{alaoui_livinglab_2013}. Seniors were part of an iterative application design process where they were encouraged to provide feedback and suggest additional functionalities to enhance existing smart TV applications. 

However, these studies remain in the realm of well-known participatory design principles. In our research we try to engage senior end users into all prototype design phases - including problem analysis and the full feature and UI design process. In \cite{ladner_design_2015} this approach is referred to as "design for empowerment".

Teaching older adults about UX design has the potential to transform their relationships with developer teams in participatory design settings. Familiarizing seniors with any other software development skills (e.g. programming) would not be as beneficiary as the key to "design for empowerment" is to retain and focus on the end user's perspective and UX design is the best place to do so. For seniors to proactively take part in designing their own applications they must first understand the conceptual model of the application prototyping process and especially UI design. In our study we examine how older adults build their understanding of user interfaces and search for a method to enable them to effectively collaborate with younger development teams.

To the best knowledge of the authors, the "design for empowerment" approach - i.e. moving from "designing with older adults" to "designing by older adults" - has not yet been investigated.

\section{Course settings}

\subsection{Settings}

This study analyzes an entry level application prototyping course for older adults consisting of two 4-hour modules. The goal of the course was to provide elderly participants with the knowledge and tools to be able to understand and take part in designing basic applications.

On the first day of the workshop the participants were introduced to basic UX concepts, i.a. types of UIs, basic UI elements (screens, text, images, buttons, links, text entry fields), interactions (clicks, scrolling, text entry, gestures), usability, accessibility, aesthetics, preparing paper prototypes. They were then divided into teams of 2 or 3 and asked to create a paper prototype for a simple application given an outline of its functionalities. Afterwards, the participants were shuffled between groups and tested each other’s prototypes.

The second module was set in a computer laboratory and introduced the participants to Adobe Experience Design (Adobe XD) - a prototyping tool for preparing interactive UI mockups. After learning the basics of Adobe XD (creating a new project, adding screens, shapes, buttons, text and simple click interactions), the participants were asked to individually create interactive mockups for the application they paper-prototyped on the previous day. They had access to the tutor and were able to ask questions at any time.

The workshop was conducted for two different groups of seniors. The first group consisted of technology-aware elderly participants who regularly use PCs or mobile devices. The second iteration of the course was aimed at older adults with little or no prior ICT experience.

Based on the results of the two workshops a methodology for teaching UX and prototyping to seniors shall be proposed and evaluated on a third group of participants in a consecutive study.

\subsection{Participants}

The course participants were recruited among seniors organized around a Polish society dedicated to helping older adults expand their knowledge and give back to society (Stowarzyszenie Kreatywni 50+). The participants were recruited to the two course groups according to their general technical proficiency. Those with little or no experience with ICT were selected for the beginners group while those who declared using technology regularly were admitted to the advanced group. Before the start of the workshop the participants were asked to fill in the DigiComp questionnaire \cite{digicomp} to verify their declared computer skills. The skill averages calculated for the two groups confirmed that the group attribution based on participants' subjective skill declarations proved successful, i.e. the advanced group achieved significantly higher skill index values than the beginners.

The beginners group consisted of 9 participants (7 female, 2 male) aged between 60 and 80 years old (the median age was 69, average age: 68.6) while the advanced group consisted of 15 people (7 female, 8 male) aged between 62 and 88 (median: 71, average: 70.8). 

\section{Methodology}

A three component methodology was used to assess the results of the workshop. The three qualitative research tools used are: self-assessment questionnaires filled in by participants, in-class observation by the research team and post workshop interviews - both individual and affinity group interviews \cite{hennink_2007}.

At the start of the course, participants were asked to fill in a basic demographics survey, the standard Future Time Perspective questionnaire and the DigiComp digital competencies evaluation questionnaire \cite{digicomp}.

During the course the research team was present in the classroom and observed the participants’ progress and their learning process. An observation sheet was provided but taking additional notes was encouraged.

At the end of each module (day), the participants filled in an exit questionnaire which included questions concerning the perceived difficulty of the module, their interest in the topic and willingness to learn more about it.

Four weeks after the course, post-workshop interviews were conducted with the participants to verify whether the course had any lasting impact on their interest in designing new technologies and whether they have internalized the general principles of UI design. The interviews (affinity group and individual) were conducted by a trained psychologist from the research team.

\section{Results}

\subsection{Observations}

\subsubsection{Beginners group}

During the first day of the workshop (introduction to UX and prototyping) the participants of the beginners group were generally quiet and did not interact much with the lecturer. Their understanding of technologies was very limited - when asked whether they owned smartphones one of the seniors held up his device asking whether it was a smartphone or not; when shown a video presenting VR solutions another asked whether a VR coffee machine makes real coffee; several participants were not familiar with the QWERTY keyboard layout; no one reported ever using an ATM or ticket vending machine.

During the practical exercise (preparing a paper prototype in small teams) the participants read the problem statement (a short description of a simple ATM machine) very carefully before proceeding further. Initially, they were not sure how to start. All of the teams began by analyzing the requirements and taking notes. One team realized after a while that they were in fact rewriting the problem statement instead of designing screens for a solution. The other two groups needed a hint from the tutor before they realized that they should be designing actual screens to be presented on the ATM machine.

The participants' lack of technical awareness showed numerous times, e.g. they did not know what a PIN code was. Nevertheless, they first tried to recreate an ATM interface based on what little knowledge they had about real ATMs (e.g. from observing younger people using them) and only later tried to design their own solutions. All of the teams considered basic usability issues they saw as relevant ("it should say which way to put the card in" remarked one participant, one group proposed a UI which asked the user to select specific banknotes to be withdrawn instead of just choosing a total sum of money).

All of the groups paid much attention to the wording used in the interface and debated it thoroughly. Two participants insisted on the interface being "polite" and thanking the user for using it ("what about manners?" they exclaimed). 

In the prototype testing phase, the designers of each solution were unwilling to allow the tester (a participant from another group) to use their prototype freely - despite being instructed on the importance of doing so ("don't mind me" said one participant to a tester after clicking through the whole prototype himself). All of the teams were irritated when the tester did not understand their UI and no-one decided to redesign major aspects of their solutions after seeing them fail in the testing phase - most teams just fixed minor issues (e.g. unclear text messages, button placement).

During the second module participants prepared interactive mockups in Adobe XD. Most of the seniors were not proficient in using desktop computers ("I would have drawn this quicker by hand" exclaimed one) and initially did not consider Adobe XD easy to use ("This program is not intuitive - it's badly designed!" one senior said) but with some help and repetition all managed to follow the tutors instructions and learned to add screens, simple UI elements and interactions. The two most technically proficient seniors tried to explore the program on their own. One of them was initially reluctant to ask for help when lost. After a while the seniors started helping each other and seemed more and more engaged. Some were even hesitant to leave the laboratory for the coffee break.

All of the seniors managed to produce working interactive mockups. One corrected errors in the interface he designed on the previous day before transferring the prototype to the computer, others merely redrew their UIs in Adobe XD. At this stage the seniors expressed their interest and engagement ("this is so fun!" exclaimed one, "can I print my interface to show my family?" asked others, some took photos of themselves with their mockups).

The UIs drawn by the participants were very basic but many of the seniors gave much attention to detail (font, button colors). Most did not make use of the whole screen and placed all the UI elements in its center. UI elements and font sizes were relatively large and thus quite accessible but buttons were often close together which would result in misclicks if the UI was implemented. Moreover, when accessibility (e.g. large fonts) got in the way of the interface's "politeness" (e.g. long texts displayed to the user) the latter was considered a priority. 

Even though the seniors were informed that they can view and test their UIs in real time on tablets connected to the computers, none of them did so before completing the mockup. Afterwards, several participants briefly tested one or two test scenarios using the tablets but others just viewed their prototypes.

\subsubsection{Advanced group}

Contrary to the beginners, the advanced group was very active during the first day of the workshop. 5 participants diligently took notes during the lecture, one recorded large portions of it using her tablet, another took photos of the presentation slides. Primarily male participants contributed to the discussion while women were more likely to take notes.

When asked about their understanding of UIs two of the most active participants replied that UIs were a method of communication between man and machine. Other participants were unable to produce a definition and asked whether ticket vending machines, icons displayed on railway station displays and even car keys qualify as UIs. On the other hand, when presented with a tangible UI example i.e. a door knob, the majority of the group took part in discussing the merits and drawbacks of various solutions for the door user. While discussing other everyday UI examples the seniors stressed that the speed of change is a key factor in usability. They also recalled their attempts to use ATMs, parking meters or ticket vending machines and pointed out that even though they generally understood the UIs they encountered, they got lost in the process of using them and needed to ask for help or repeat the operation.

Most of the seniors used terms such as “widget” and “link” fluently. Some of them shared their experiences with early computers during the discussion. On the other hand, when a touchpad was mentioned in the course of the lecture, the seniors asked for an explanation of the term - only to find that they knew the device but were not aware of its name. Similarly, when icons were mentioned (in a mobile app context), the participants asked for various icon meanings and eagerly took notes. When discussing UI elements, some seniors showed the widgets being spoken about to their less active colleagues on their smartphones.

During the paper prototype exercise, the seniors had difficulty in understanding what a paper prototype was and why it is important to produce one before implementing the whole solution. Initially, the teams started by discussing the problem statements and all of them tried to generalize their solution to include various (sometimes quite obscure) scenarios. When encouraged to take into account only a small number of the most likely scenarios, the groups in turn focused on technical details and feasibility of the solution instead of simply proposing a UI. Though instructed to assume the backend works and focus on the UI, problems such as finding the best methods for connecting devices and scanning product codes were discussed. Even after refocusing their attention to designing just the frontend, the participants still regularly got caught up in discussions about technical solutions needed to provide the desired functionality. 

Only one group started by drawing a start screen for their application. The approaches taken by other teams ranged from a linear, scroll-through UI for the whole application and widgetless screens with simple, centered text commands for the user to a flow diagram outline for the whole application. The use of colors and shapes in the prototypes was quite random. None but one of the teams introduced their own design convention (deciding buttons shall be represented by sticky-notes of a specific color, system feedback by another etc.) Some teams did not even reach for the colored pens, paper of sticky notes available.

Appropriate wording and icons were, however, an important part of the teams’ discussions. The seniors often recalled solutions or functionalities they knew from their own use of PCs and the Internet and tried to incorporate them while designing the UIs. One group also discussed accessibility issues and provided solution ideas for users with various disabilities. 

By the end of the first day the groups were at very different stages of their paper prototype task. Two groups completed their prototypes and managed to test them, the others barely finished their drawings. In most teams two of the participants were actively engaged in drawing the prototype while the third only consulted their solutions.

As in the beginners group, when testing the prototypes, the seniors generally did not allow the tester much freedom. Most in fact performed a presentation of their UI, explaining thoroughly not only how to use the prototype but also the underlying technical solutions. The testers eagerly accepted this and gave their comments on the UI as if they were team members and not potential users. This sometimes turned the testing process into a heated debate between both sides.

During the second module, the process of learning Adobe XD is quite smooth. Most issues arise when participants experiment with the software on their own (which they do eagerly). They were willing to help each other and try to solve problems on their own when possible but were also open to asking questions.

By the end of the workshop all of the seniors managed to produce at least two screens connected by a “click” interaction though barely anyone managed to transfer their whole paper prototype to the mockup. The seniors put much effort into rearranging and coloring their UI elements. One senior was thrilled by the possibility of connecting two screens with a “click” interaction which enabled him to move back and forth between them and proudly presented his discovery. Several participants stayed after the workshop to finish their mockups and asked about the possibility of obtaining Adobe XD to use on their home PC. However, at least one participant seemed to view Adobe XD as a graphics editor rather than a prototyping tool even when shown the working mockup.

\subsection{Participants’ evaluation}

Participants completed an evaluation questionnaire after each module of the workshop. The questionnaire used a 7 point scale where 1 was adequate to "completely disagree" and 7 to "completely agree".

Results suggest that both groups found learning UX design interesting (14 out of the 24 participants gave the highest mark and the overall score was on average 6.5 out of 7) and claimed they would gladly learn more (16 max scores, an average of 6.8). Most seniors also highly rated the courses "usefulness" (14 max scores, an average of 5.76) and declared they would like to put their newly acquired skills into practice (9 max scores, an average of 5.96). The practical tasks (paper prototype and interactive mockup) were deemed moderately difficult by both groups with average scores of 4.4 and 3.6 respectively (advanced group) and 4.9 and 4.4 (beginners group) where 1 was adequate to "very easy" and 7 to "very difficult" - though these scores varied greatly between participants.

\subsection{Learning outcomes}

After the first workshop module all teams in both groups produced a form of paper prototype, though the quality of the solutions varied considerably. The advanced group (which was only shown a sample paper prototype during the lecture and given no additional guidelines concerning the expected format of their solution) came up with various different strategies for drawing their prototypes. Only one advanced team took a typical UX design approach and prepared a set of screens which could be consecutively shown to the user to form a paper prototype. Another team took a more interaction based approach and drew a flow diagram of the user's interactions with the application without actually designing the specific screens. A third team drew their prototype without dividing it into separate screens - ending up with a long, scrollable page which guided the user through a single use case. Other advanced teams produced screen-like solutions but initially used mainly text messages to guide the user through their interface only adding buttons and other UI elements after being hinted to do so.

Almost all the teams found it difficult to decide on the design conventions to use for their prototypes. Only one group used specific colors of sticky-notes to depict specific UI elements (e.g. pink sticky-noted were reserved to buttons, green ones contained system feedback) - all the others did not provide much consistency in their prototypes which resulted in the creation of confusing UIs where buttons, system texts and text entry fields looked alike.

The beginners group received more guidance on the expected form and look of the paper prototype (they were shown several sample paper prototypes during the lecture and saw a live paper prototype demonstration by the tutor). This group generally reproduced the suggested format though again not all of the prototypes were clearly divided into separate screens and the participants also showed a tendency to predominantly use text instead of graphic UI elements.

When working on the prototypes, in most teams only 2 of the 3 group members actively took part in designing the prototype. The third person usually took the role of an observer and occasionally commented on what was being done. In mixed-sex groups the observer was usually female.

On the second day several seniors from both groups reported being more aware of UIs in their day to day lives. They gave examples of such interfaces including ticket machines, ATMs and parking meters and reported to have looked at them more carefully or intentionally watched others using them.

By the end of the workshop everyone from the beginners group had managed to transfer their paper prototypes to Adobe XD and produce a simple mockup. In the advanced group, most participants did not manage to do this, though most did have several working, often quite detailed, screens to show. Seniors in the advanced group put too much effort into creating an actual, working application instead of just supplying its UI which is why they did not manage to finish their mockups on time.

\subsection{Post-workshop interviews}

Five weeks after the workshop, individual and group interviews with selected participants were conducted to verify whether the workshop had any lasting impact on their understanding of UI design.

When asked to define a user interface, just several seniors from the advanced group produced a satisfying definition (i.e. one pointing to UIs role in human-computer interaction). Less advanced participants were initially unable to verbalize a definition but when encouraged using a visual based method (they were given visual and verbal prompts to help them express their opinions by choosing and describing pictures), they described UIs as "methods of communication between people" or "instructions on how to use an appliance" - both of which suggest they did not fully grasp the concept of a UI though they did understand some of its key purposes. The interviewees were also unaware of the special place prototyping and UX hold in ICT - they saw it as one more technology to be learned without realizing its wider implications. Interestingly, one of the participants (from the beginners group) thought UI design was synonymous to programming.

On the other hand, all of the seniors emphasized that a good UI should be simple, easy to use and intuitive. They noted that bad UIs are common and that some negative experiences with technology may result from flawed UIs and not the users' lack of technical knowledge.

When asked about the traits of a good UX designer, most participants claimed he (just one envisioned a female) should be young, energetic, tech-savvy and artistically talented. On exploring this idea further, they claimed that even though they consider themselves above averagely active for their age group, they still see an wide gap between their comprehension of technology and that of the younger generation. Comments such as "all young people can be good UX designers because they are so used to technologies" were uttered. Such opinions suggest the seniors had stereotypes about UX designers, viewing them as people who greatly differ from themselves. These stereotypes may have a strong impact on their motivation for learning UX since, according to self-efficacy theory \cite{bandura_analysis}, people are more motivated to try new activities when they see that others similar to them are successful in the activity. This may be an issue worth addressing as this lack of self-confidence influenced the senior's outlook on their potential role in an application building team: none but one declared they would like to take responsibility for prototyping a real application, even in a workshop or hackathon setting. Even participants from the advanced group claimed they prefer to remain testers or consultants for younger designers' ideas.

\section{Conclusion}

Observing older adults' process of understanding UI design principles has led us to a better comprehension of the models they use when interacting with technology. Key findings include: difficulties in separating frontend from backend and limiting the scope of the designed solution, and using a linear, storytelling-like approach to UI design instead of a screen-centered approach.

Tech-savvy seniors were deeply inclined to build fully-functional applications and focused on implementation details (e.g. determining the technologies to be used, providing reasoning for the feasibility of a solution). They did eventually refocus their attention towards designing only the UI as they had been instructed to but questions about technical aspects often reappeared in their discussions. Novice technology users were slightly less prone to dwelling on aspects of the backend of their prototypes, however, understanding the distinct division between UI and backend and the possibility of designing them separately remained a major issue. 

Both groups also gave much attention to special (and often obscure) use cases while analyzing the problem statement and had difficulty in prioritizing usage scenarios and limiting the scope of their solutions to address only the most important use cases. This tendency was exceptionally prominent in the advanced group (though beginners also demonstrated its symptoms) and led to the creation of overly complex UIs. Conversely, the seniors did not consciously look for corner cases or pay much attention to error handling (e.g. providing a way for the user to undo an action). 

An interesting observation was that the participants generally produced linear, text-based prototypes resembling the transcription of a storytelling session rather than screen-based, UI-oriented solutions. This tendency was stronger in the beginners group though the advanced group also struggled with accepting screens as the basic building blocks of their prototypes. This observation is most surprising in the context of the advanced group who's members were regular PC and mobile device users and could have been expected to have built up an intuition for screen-based application design as a result of their everyday exposure to technology.

In the light of our findings, when working towards creating an effective methodology for teaching older adults abstract technology skills (e.g. UI design), aspects such as abstract thinking abilities, motivation, self-confidence and the presence of a role model should be addressed.

We suggest starting the workshop with abstract thinking exercises (e.g. drawing symbols for abstract terms). Another suggestion is to provide a positive role model by showing examples of older adults actually performing UI design (e.g. showing photos of senior UX designers during the lecture) as this could enhance the participants' motivation and self-confidence. Additional motivation could be achieved by providing real-life examples of participatory design success stories which could help the seniors feel empowered to create applications needed by their communities and take responsibility for designing a real application. It might also be useful to stress the important role UX designers play in society since they help people better communicate with the modern, digital world (a growing number of studies indicates that seniors involved in productive, socially important activities have greater life satisfaction \cite{vanwillingen_differential}). Thus, we suggest that UX designing be presented as an attractive opportunity for helping society.

Participants should also be shown widely-used conventions used in UI design and provided with guidelines and instructions on where to start and how to conduct the design process step by step to help them structure their work and let them feel more secure when presenting it to younger colleagues.

Future research on this subject shall include testing the proposed guidelines in a setting where seniors are put in the role of UX designers and cooperate with programmers to prepare a working application. Additional analyses of older adults conceptual models of technology should also be conducted. Afterwards, further steps can be taken to identify and solve other issues faced by seniors in the context of participatory design to facilitate the process and take full advantage of their user perspective.

\section*{Acknowledgments}

This project has received funding from the European Union’s Horizon 2020 research and innovation programme under the Marie Skłodowska-Curie grant agreement No 690962.

\bibliographystyle{splncs03}
\bibliography{references.bib}
%
%

\end{document}